\documentclass[sigconf]{acmart}
\setcopyright{none}                       
\renewcommand\footnotetextcopyrightpermission[1]{}  
\AtBeginDocument{%
  }
\usepackage{graphicx}
\usepackage{algorithm,algorithmic}
\usepackage{enumitem}
\usepackage{adjustbox}

\definecolor{salmon}{RGB}{250,128,114}
\newcommand{\jz}[1]{\textcolor{black}{#1}}

\copyrightyear{2026}
\acmYear{2026}
\setcopyright{cc}
\setcctype{by-nc-nd}
\acmConference[DAC '26]{63rd ACM/IEEE Design Automation Conference}{July 26--29, 2026}{Long Beach, CA, USA}
\acmBooktitle{63rd ACM/IEEE Design Automation Conference (DAC '26), July 26--29, 2026, Long Beach, CA, USA}
\acmDOI{10.1145/3770743.3804091}
\acmISBN{979-8-4007-2254-7/2026/07}

\begin{document}

\newcommand{\M}{PHAROS}
\title{\M: Pipelined Heterogeneous Accelerators for Real-time Safety-critical Systems With Deadline Compliance
}

\settopmatter{authorsperrow=3}

\author{Shixin Ji}
\affiliation{%
  \institution{Brown University 
  \city{Providence}
  \country{USA}
  }}
\email{shixin_ji@brown.edu}

\author{Jinming Zhuang}
\affiliation{%
  \institution{Brown University 
  \city{Providence}
  \country{USA}
  }}
\email{jinming_zhuang@brown.edu}

\author{Sarah Schultz}
\affiliation{%
  \institution{Brown University 
  \city{Providence}
  \country{USA}
  }}
\email{sarah_schultz2@brown.edu}

\author{Zhuoping Yang}
\affiliation{%
  \institution{Brown University 
  \city{Providence}
  \country{USA}
  }}
\email{zhuoping_yang@brown.edu}

\author{Xingzhen Chen}
\affiliation{%
  \institution{Brown University 
  \city{Providence}
  \country{USA}
  }}
\email{xingzhen_chen@brown.edu}

\author{Zheng Dong}
\affiliation{%
  \institution{Wayne State University
  \city{Detroit}
  \country{USA}
  }}
\email{dong@wayne.edu}

\author{Alex K. Jones}
\affiliation{%
  \institution{Syracuse University
  \city{Syracuse}
  \country{USA}
  }}
\email{akj@syr.edu}

\author{Yihui Ren}
\affiliation{%
  \institution{Brookhaven National Laboratory
  \city{Upton}
  \country{USA}
  }}
\email{yren@bnl.gov}

\author{Peipei Zhou}
\affiliation{%
  \institution{Brown University 
  \city{Providence}
  \country{USA}
  }}
\email{peipei_zhou@brown.edu}

\renewcommand{\shortauthors}{Shixin Ji, Jinming Zhuang, Sarah Schultz, Zhuoping Yang, Xingzhen Chen,\\ Zheng Dong, Alex K. Jones, Yihui Ren, and Peipei Zhou}

\begin{abstract}
Spatially partitioned heterogeneous accelerators (HAs) are increasingly adopted in embedded systems for their performance and flexibility. 
Yet most existing HA design frameworks optimize primarily for throughput or quality-of-service (QoS) metrics.
They often overlook safety-critical real-time requirements, including hardware support for predictable execution, real-time-aware design space exploration (DSE), and rigorous schedulability analysis. 
These requirements are essential in safety-critical applications such as smart transportation, where schedulability guarantees directly affect system safety.
To address this gap, we present \M, a real-time-centric HA design framework. \M~introduces preemption mechanisms and scheduler designs for spatially partitioned HAs under first-in-first-out (FIFO) and earliest-deadline-first (EDF) policies. Leveraging modern real-time theory, we further develop a soft real-time (SRT) schedulability-oriented DSE with objectives and constraints tailored to SRT-schedulability. 
Through comprehensive modeling, analysis, and evaluation across diverse applications, we show that \M’s schedulability-oriented DSE discovers more feasible configurations for a broader range of task sets than throughput-oriented DSE baselines, while delivering improved real-time performance. We also provide response-time analyses for the supported scheduling algorithms.
\end{abstract}

\maketitle

\vspace{-7pt}
\section{Introduction}
\label{sec: introduction}
\vspace{-3pt}

\begingroup
\renewcommand\thefootnote{}
\footnotetext{This paper has been accepted at DAC 2026}
\endgroup

Heterogeneous accelerators (HAs) are becoming increasingly popular in embedded systems due to their ability to spatially partition compute resources for different workloads, enabling higher performance and better energy efficiency~\cite{kwon2021herald,dally2020domain,zhuang2023charm,choi2020prema}. 
However, most existing HA designs focus on throughput or QoS metrics and overlook the requirements of real-time safety-critical systems, which demand predictable hardware behavior, real-time-aware design space exploration (DSE), and rigorous schedulability analysis. 
Missing a deadline can directly violate system-level safety guarantees (hard real-time, HRT), or cause severe degradation (soft real-time, SRT).

Designing real-time-oriented HAs therefore introduces a design space fundamentally different from throughput-oriented approaches. 
The primary requirement is schedulability.
In the SRT setting considered in this work, every job must complete within finite response time.
Traditional performance metrics are inadequate as optimization objectives because they neither quantify schedulability nor differentiate between feasible SRT designs.
Moreover, dynamic scheduling and preemption mechanisms require hardware support, whose timing behaviors must be modeled and analyzed. Existing frameworks thus fall short for SRT-oriented DSE.

\begin{figure}
    \centering
    \includegraphics[width=0.87\linewidth]{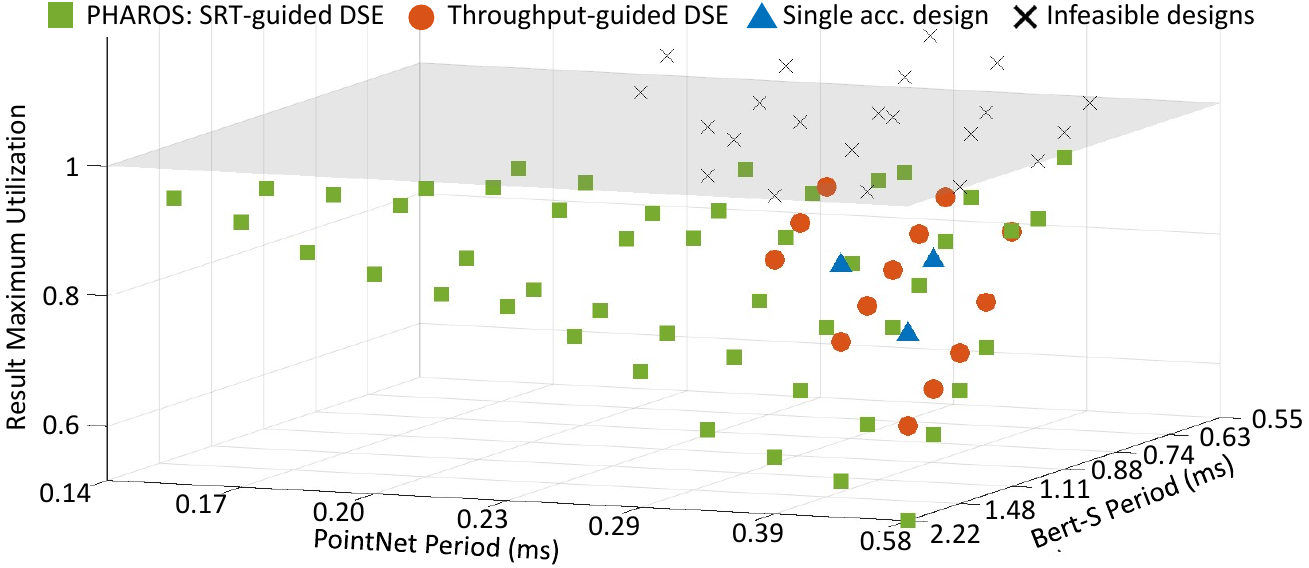}
    \caption{Comparing SRT-schedulability of different DSE results on PointNet \& Bert-S blocks combination.
    }
    \label{fig: motivation_example}
    \vspace{-20pt}
\end{figure}

Figure~\ref{fig: motivation_example} illustrates this challenge: we compose different tasksets via assigning different task periods to the same application combination of PointNet and Bert-B block, then create HA systems using different DSE methodologies and measure their SRT schedulability.
Smaller period means heavier workload.
Each point on the figure represents a generated HA system.
We report the maximum utilization(execution time/ period) among all accelerators within each HA system. When this maximum utilization $>$ 1, there is no way for the system to be SRT-schedulable.
In this example, only three tasksets are SRT-schedulable under a fixed accelerator design. 
Throughput-guided DSE can customize hardware to make 13 tasksets schedulable, but still misses most feasible configurations.

To address this gap, we propose \M, a pipelined HA architecture tailored for SRT-schedulability. Guided by state-of-the-art real-time theory~\cite{dong2017optimal}, 
\M~enforces the utilization of each accelerator to remain below one,
and adopts a pipelined execution model where tasks must complete all segments on one accelerator before moving to the next. 
Based on these constraints, we formulate an SRT-oriented DSE, selecting maximum utilization as the optimization objective because it reflects hardware efficiency while aligning with SRT schedulability. 
We further introduce a beam search heuristic to efficiently explore this distinct design space.

As foundational support for the DSE, we implement hardware mechanisms for preemption and dynamic scheduling (FIFO and EDF). 
The system is modeled in a fully preemptive context, accounting for hardware latency and preemption overhead to ensure safe, accurate timing analysis. 
As shown in Figure~\ref{fig: motivation_example}, \M~discovers feasible designs for 49 tasksets and reaches a 3.76× increase over throughput-guided DSE. We evaluate \M~on diverse real-world workloads. 
\M~finds 1.44$\times$ to 2.28$\times$ more feasible solutions and achieves up to 6.2\% reduction in maximum utilization among the feasible solutions.
We also analyze response times under different scheduling policies and discuss their tradeoffs. 
In summary, our contributions are as follows:
\vspace{-3pt}
\begin{itemize}[leftmargin=*]
\item \textbf{SRT-oriented HA design space exploration:} 
With respect to the guideline real-time theory, we formulate the~\M~DSE.
This SRT-oriented DSE is distinct from existing throughput- or QoS-oriented ones in constraints, objectives, and strategies.
We propose a beam search heuristic with near-optimal quality.
\item \textbf{Preemption and dynamic scheduling hardware support for HAs:} 
We design the preemption mechanism and dynamic scheduler to \M~system, which are critical to SRT systems.
\item  \textbf{Comprehensive modeling and analysis:} 
We model \M~in a fully-preemptive real-time system scenario, handling the hardware latency and preemption overhead.
We also provide response time analysis for the different scheduling methods provided.

\end{itemize}

\vspace{-10pt}
\section{Related Works}
\label{sec: related works}

\noindent\textbf{Hardware accelerators for real-time safety-critical systems:~}HA designing has been explored by various works on different platforms\cite{
zhuang2023charm,
ssr,
kwon2021herald,
basalama2025streamhls,   
li2023fixed, 
eq_vit_tcad, 
yang2023aim,
fpga25aries,
zhang2020dnnexplorer,
cai2023set}.
However, existing frameworks remain primarily throughput- or latency-oriented in their accelerator designs and typically rely on static or FIFO scheduling policies that lack preemption. Consequently, they cannot be directly deployed in SRT safety-critical systems, where heterogeneous workloads must be scheduled according to differing levels of importance (i.e., task priorities).

There are also frameworks implementing real-time system-related features such as preemption or deadline-aware scheduling \cite{Dream,choi2020prema,ghodrati2020planaria,wang2023cd,gao2023layer,oh2021layerweaver,kim2023moca,zeng2022serving}.
These frameworks primarily optimize for QoS metrics like 99\% latency, which allows some jobs to not respond.
Unlike these metrics, SRT deadline compliance requires that every job have a limited response time.
As a result, these frameworks cannot be deployed onto the SRT safety-critical systems, either.
DREAM~\cite{Dream} and CD-MSA~\cite{wang2023cd} choose to drop jobs to make room for others. 
Although QoS metrics could be improved, this is a direct violation of the SRT requirement.
Moreover, these frameworks do not test whether the system can ensure deadline compliance mathematically.

Targeting real-time systems, some frameworks are proposed to meet the deadline compliance requirements and provide analysis of schedulability or response time.
\cite{restuccia2021time} implements a non-preemptive DNN-based driving assistance system based on the Vitis-AI~\cite{vitisai} framework.
MESC~\cite{guan2024mesc}, ART~\cite{ji2025towards,ji2025art}, and DERCA~\cite{DERCA} implement the preemption mechanism and EDF scheduling.
These works are tailored for real-time safety-critical systems; however, their performance is limited by the single-accelerator architecture.
The hardware customization is thus limited.
As a result, \M~can ensure deadline compliance for tasksets with smaller periods, whereas these frameworks fail due to hardware inefficiency.

\noindent\textbf{Real-time theories for guiding accelerator design:~} 
To ensure deadline compliance, the accelerator design must be paired with real-time theories to analyze schedulability or response time.
For a single accelerator, ~\cite{liu1973scheduling} gives the earliest formulation in the HRT scenario, where every job is required to be finished before its deadline.
~\cite{devi2005tardiness} analyzes the SRT scenario, where the jobs are only required to have limited response times.
For multiple heterogeneous accelerator systems, ~\cite{jiang2023blueface} and ~\cite{dong2022schedulability} provide analyses under HRT and SRT scenarios, respectively, with no constraints placed on the system.
However, considering many corner cases, these two works are overly pessimistic, making the candidates that pass the schedulability test inefficient in hardware.
\cite{dong2017optimal} gives a tighter bound on the SRT scheduling by constraining the topology of the accelerators, that is, a job must finish all its execution on one accelerator before proceeding to another, and no backtracking is allowed.

We choose to leverage \cite{dong2017optimal} to guide~\M. 
We combine the hardware preemption operation into the worst-case execution time (WCET) to model the system in a fully-preemptive scenario, and design our DSE with respect to the constraints.
We acknowledge that this existing theory constrains design space; still, it can cover a huge part of spatial accelerator designs.
We leave the exploration of more advanced real-time theory and topologies for future work.



\section{\M~Hardware Architecture Design and Task Modeling}
\label{sec: hardware}

In this section, we introduce the proposed \M~hardware architecture, including the dataflow and control flow.
We then formulate the whole system in the fully-preemptive preemption scenario and explain our method for preemption and task model.

\vspace{-5pt}
\subsection{\M~Accelerator Design}
\begin{figure}
    \centering
    \includegraphics[width=0.9\linewidth]{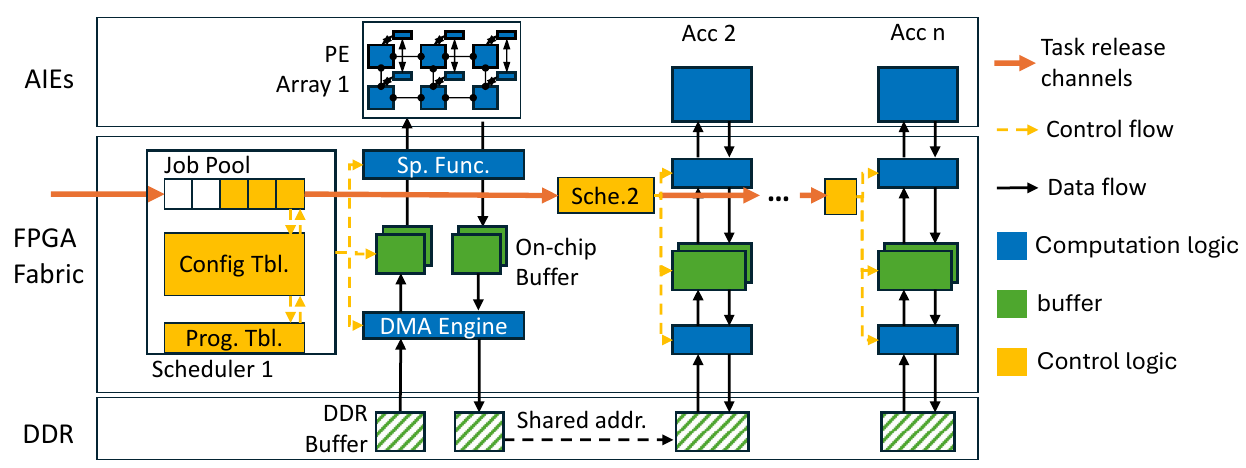}
    \vspace{-12pt}
    \caption{\M~hardware architecture.}
    \label{fig: hardware_architecture}
    \vspace{-15pt}
\end{figure}

\M~leverages a widely used accelerator model, as described in Figure~\ref{fig: hardware_architecture}, which consists of multiple independent dataflow accelerators.
In each accelerator, the execution of every DNN layer begins with loading data from off-chip DDR memory and ends with storing results to DDR. 
We adopt output-stationary dataflow with tiling for each accelerator to reduce off-chip memory access.
\jz{We apply double buffering to the set of input and output on-chip buffers to overlap the load and store with computation.}

To handle layers with various shapes in one accelerator, we store the configuration of each layer in the corresponding on-chip scheduler, including the shape (e.g., M, K, N in matrix multiplication) and the DDR base address.
\jz{The partitioning of DDR memory space is determined before compilation time.}
We allocate separate \jz{memory space} for different tensors to avoid conflict.
\jz{For intermediate data that is produced and consumed by consecutive layers, we automatically maintain a unified buffer space to avoid duplicate memory consumption.}
\jz{We design dedicated DMA engines to issue AXI transactions to off-chip DDR memory, enabling interleaved access that allows multiple accelerators to share the available bandwidth.}
\vspace{-10pt}
\subsection{Control Flow and Scheduling Methods }
\jz{To ensure correct and efficient execution, a scheduling mechanism is essential in determining the execution order of each layer on every accelerator and in synchronizing data across accelerators.}
In \M~we apply a decentralized control flow with each accelerator scheduling independently.
As shown in Figure~\ref{fig: hardware_architecture}, each on-chip scheduler has a static configuration table, a dynamic progress table to store the execution status, and a job pool to store the ready jobs that can be scheduled.
When FIFO scheduling is used, the job pool is implemented as a queue. When EDF is used, it is implemented as an array sorted by deadline before scheduling.
If the new job's deadline is smaller than the ongoing one, preemption happens.

The maximum capacity of the scheduler tables is fixed at compile time.
We set the maximum entry of the progress table and job pool to the number of tasks in the taskset, because when the system is SRT-schedulable and accelerators are in a chained topology, at most one ready job from each task exists on an accelerator.
The time and resource overhead of the schedulers is small, with each entry taking $<$25 bytes of data and $<$1 µs sorting time for the arrays.

\M~connects the job pool between different accelerators via FIFO queues implemented by HLS streams. 
For the first accelerator, the jobs are released from \jz{external devices}, like CPUs or IOs connected to sensors.
Each accelerator will release one job to its successor after finishing the segment on this job.
This means for one job, the segment on one accelerator is not ready unless (1) this is the first accelerator or (2) all predecessor accelerators finish execution, which is compatible with the guideline theory~\cite{dong2017optimal}.

\vspace{-10pt}
\subsection{Task and Performance Modeling}
Following the setup in \cite{dong2017optimal}, we first formulate the \M~system under a fully preemptive scheduling model and then discuss how hardware-level preemption latencies influence its behavior. 
A task set $\tau$ is executed, where each task $\tau_i$ is characterized by tuple $(e_i, p_i, d_i)$. 
Here, $e_i$ denotes WCET, $p_i$ is the release period for periodic tasks (or the minimum inter-arrival time for sporadic tasks), and $d_i$ is the deadline. In this paper, we assume an implicit-deadline model with $d_i = p_i$. An SRT-schedulable real-time task system allows deadline misses but requires that every task’s response time remain bounded. 
During runtime, each task $\tau_i$ generates a sequence of jobs with identical inference shapes, denoted by $\tau_{i,j}$.

To execute the task set, we model the system as a pipeline of $M$ accelerators, denoted $acc^1, \ldots, acc^M$. 
Each task $\tau_i$ is correspondingly decomposed into up to $M$ segments, where $\tau_i^k$ denotes the segment of task $i$ executed on accelerator $acc^k$, and $e_i^k$ is its WCET. 
We assume that each task is (or can be topologically sorted into) a sequence of DNN layers. 
We discuss the limitation of this assumption in Section~\ref{sec: discussion}. 
We denote the $j$-th layer of task $i$ by $l_{i,j}$, and the total number of layers in task $i$ by $L_i$. 
Layers can be mapped onto accelerators in various ways, subject to the following pipelined-topology constraint: a layer $l_{i,j}$ may be mapped to accelerator $acc^k$ if and only if either (1) $j = 1$, or (2) all preceding layers $l_{i,m}$ with $m < j$ are mapped to accelerators $acc^n$ with $n \le k$. 
We also allow an accelerator to host zero layers of a task, in which case that task simply bypasses the corresponding accelerator.

Each accelerator is characterized by several key design parameters.
Following CHARM~\cite{zhuang2023charm}, we model $acc^k$ as a \jz{3D PE array of size} $A^k \times B^k \times C^k$ operating along the $M$, $K$, and $N$ dimensions of the matrix multiplication. 
The corresponding on-chip tile size is $X^k \times Y^k \times Z^k$. 
As a result, the critical hardware resources consumed by $acc^k$ can be derived, including the number of PEs/AIEs, the on-chip bandwidth between the AIE array and the FPGA fabric, the on-chip memory usage, and the off-chip DDR bandwidth.

We denote the execution latency of layer $l_{i,j}$ on accelerator $acc^k$ as $bl_{i,j}$, which can be expressed by:
\vspace{-2pt}
\begin{equation}
bl_{i,j} = Exec\big(l_{i,j}, A^k, B^k, C^k, X^k, Y^k, Z^k\big).
\end{equation}
\vspace{-5pt}

To ensure SRT schedulability, we analyze each accelerator separately.
For $acc^k$, we define its utilization $u^k$ as the sum over all tasks of the fraction of processing time on this accelerator:
\vspace{-5pt}
\begin{equation}
u^k = \sum_i \left(\frac{e_i^k}{p_i}\right).
\end{equation}
Based on~\cite{dong2017optimal}, the system is SRT-schedulable if and only if the utilization of every accelerator does not exceed~1, which holds under both FIFO and EDF scheduling:
\vspace{-5pt}
\begin{equation}
u^k \le 1,\quad k = 1,2,\ldots,M.
\label{eq: utilization_constrain}
\end{equation}
\vspace{-20pt}

\subsection{Preemption Support \& Overhead}
\begin{figure}
    \centering
    \includegraphics[width=0.8\linewidth]{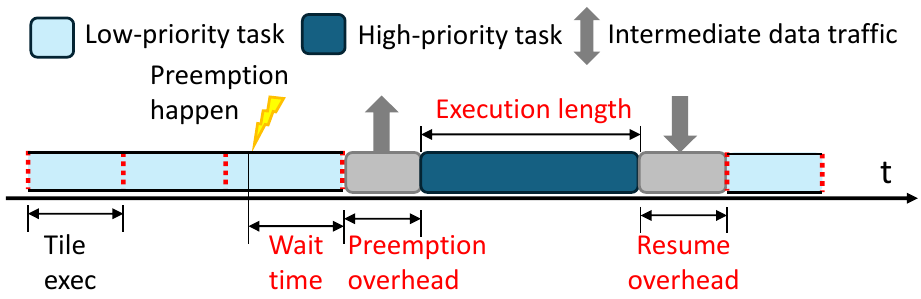}
    \vspace{-12pt}
    \caption{Preemption pattern and overhead modeling.}
    \label{fig: preemption_overhead}
    \vspace{-15pt}
\end{figure}
\jz{The original real-time theory assumes a fully preemptive scenario with zero overhead, which is unrealistic due to the latency incurred by preemption.}
We explain the preemption mechanism in \M~and the corresponding modeling as follows.

As shown in Figure~\ref{fig: preemption_overhead}, \M~supports preemption within the execution of one layer by leveraging the tiled output-stationary dataflow.
When a high-priority job arrives at the scheduler, the accelerator cannot run it immediately.
Instead, the accelerator first completes the current tile of the low-priority job.
Then it stores the partial results in the output buffer into the corresponding DDR address.
The current execution status in terms of loop iteration numbers is recorded to the on-chip progress table within the scheduler.
The accelerator then runs the high-priority job.
When the low-priority job resumes, the scheduler reloads the input and output buffers according to the loop iteration and recovers execution.

On one accelerator, one job can be preempted several times by others, but can only preempt others once it is released. 
Due to this, the WCET is modeled as two parts:
\begin{equation}
    e_i^k=b_i^k+\xi_i^k
    \label{eq: acc_execution_length}
\end{equation}
where $b_i^k$ and $\xi_i^k$ represent the execution length and overhead, respectively.
To quantify $\xi_i^k$, we assume each component in the preemption process takes the maximum.
That is, the high-priority task waits for one whole tile, then begins to store the data.
The latencies of computing a tile, storing, and loading the buffers are fixed in one accelerator and are only related to its design parameters:
\begin{equation}
    \xi_i^k = e_{tile}^k+e_{store}^k+e_{load}^k
    \label{eq: acc_preemption_overhead}
\end{equation}
Where $e_{tile}^k$,$e_{store}^k$, and $e_{load}^k$ are functions of $A^k,...,Z^k$.
Specifically, when this accelerator is skipped ($b^k_i=0$), the $e^k_i$ is also 0.
The preemption only happens when EDF scheduling is applied. For the FIFO scheduling, no preemption or overhead will happen.

\vspace{-5pt}
\section{\M~Design Space Exploration}
\label{sec: DSE}
In this section, we demonstrate the key design parameters, objectives, and constraints of the design space. 
We then propose our heuristic for reducing the search time and a complexity analysis.

\vspace{-2pt}
\subsection{\M~Design Space}
\begin{figure}
    \centering
    \includegraphics[width=0.8\linewidth]{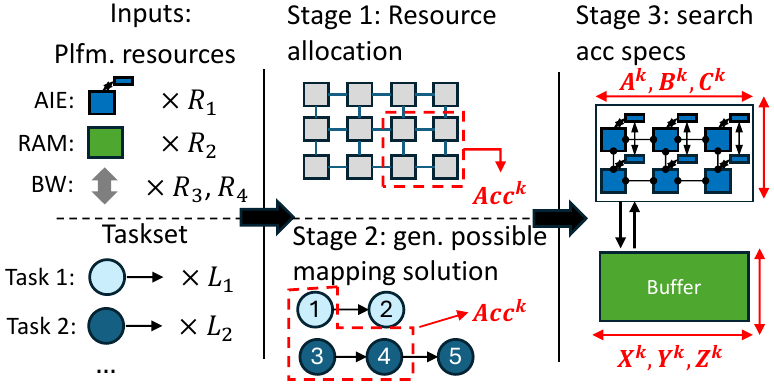}
    \vspace{-10pt}
    \caption{\M~design space exploration.}
    \label{fig: design_space_exploration}
    \vspace{-15pt}
\end{figure}

Figure~\ref{fig: design_space_exploration} demonstrates the design space of \M. 
The primary goal is find a design point to partition the platform resources to create multiple accelerators and to map the taskset to these accelerators such that the whole system is SRT-schedulable.
\jz{We explore the entire design space in three stages}. \jz{Our DSE takes as input a taskset with predetermined layer shapes $l_{i,j}$ and periods $p_i$, along with} the platform available resources $R=(R_1, R_2, R_3, R_4)$ for the AIE, on-chip memory capacity,on-chip bandwidth, and DDR bandwidth. \jz{The total number of accelerators $M$ is set as a hyperparameter during DSE.}
The first step is to partition the on-chip resources to each accelerator.
We \jz{denote} $r^i=(r^i_1,...r^i_4)$ as the resource allocated for each accelerator, with constraints $\sum_k{r^k_i}=R_i$ and $r^k_i>0$.
Next, we \jz{determine} the mapping from the layers of each task to accelerators.
\jz{Under the pipelined topology assumption, our proposed layer assignment strategy maps consecutive layers to one accelerator.}
We thus denote the mapping for $\tau_i$ as $m_i^1,m_i^2,...,m_i^M$, where $m_i^k$ denote the number of layers runs on $acc^k$, as the mapping is consecutive, the $\sum_j^{k-1}m_i^j$-th to $\sum_j^{k}m_i^j$-th layers are mapped.
All layers are guaranteed to be mapped by enforcing
$\sum_k{m^k_i}=L_i$ and ${m^k_i}\ge0$.
Finally, with the resource budget and task segments determined, we then search the $A, B, C, X, Y, Z$ to get the final design for each accelerator.
To ensure SRT-schedulability, we require all accelerators to have utilization $\leq$ 1 as shown in equation~\ref{eq: utilization_constrain}.

Instead of using conventional metrics like throughput/latency, we set the objective of \M~DSE as $\min{(\max_k{u^k})}$, the maximum utilization among all the accelerators.
This objective better quantifies the design requirements of SRT systems based on the fact that, after satisfying the utilization constraint and being SRT-schedulable, some design metrics of real-time systems can be further benefited from scaling down the task release periods.
For example, the speed of an autonomous driving system is bounded by the gap between performing two perception updates to ensure safety, thus reducing the period results in a higher maximum speed in the system.
When scaling down all periods proportionally to x\%, the utilization of all accelerators will increase to $\frac{1}{x\%}$.
Thus, the potential of one system to further scale down the periods without $u>1$ is determined by the maximum utilization of the accelerators.

\vspace{-5pt}
\subsection{Beam Search Heuristics}
One significant challenge in \M~is solving the huge design space, which grows exponentially when adding more tasks to the taskset, or applying a larger application with more layers.
To tackle these problems, we propose a beam search heuristic method.


\begin{figure}
    \centering    \includegraphics[width=0.9\linewidth]{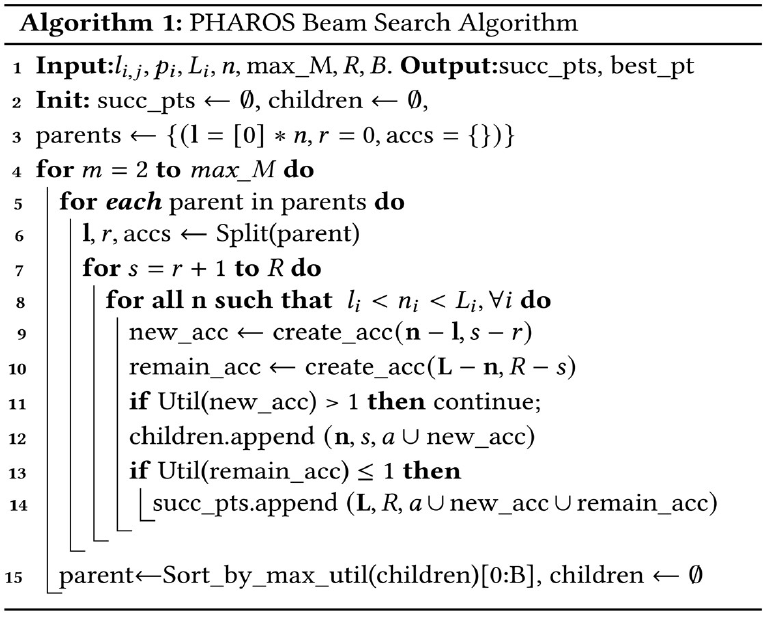}
    \label{beam_search_algorithm}
    \vspace{-14pt}
\end{figure}
\begin{figure}
    \centering
    \includegraphics[width=0.9\linewidth]{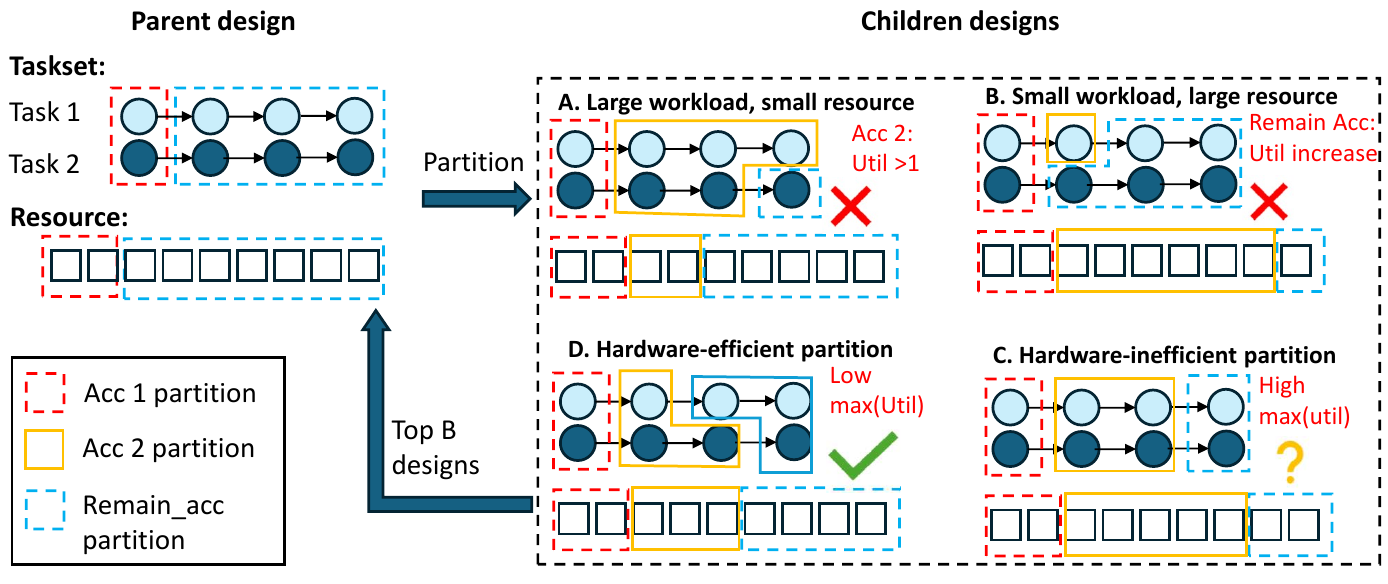}
    \vspace{-13pt}
    \caption{Beam search progress.}
    \label{fig: beam_search}
    \vspace{-12pt}
\end{figure}

\begin{figure*}
    \centering
    \includegraphics[width=0.95\linewidth]{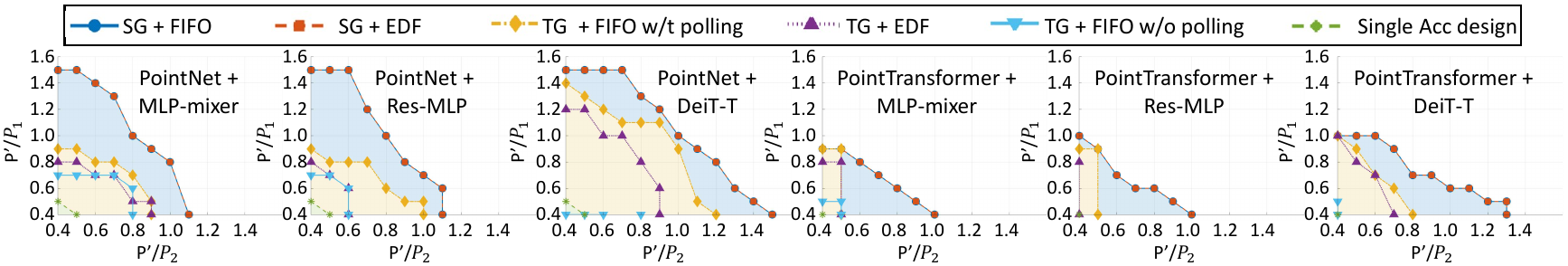}
    \vspace{-10pt}
    \caption{SRT-schedulable taskset design points of SRT-guided (SG) and throughput-guided (TG) EDF using FIFO and EDF scheduling on various application and period setups.}
    \label{fig: success_rate}
    \vspace{-10pt}
\end{figure*}
Algorithm 1 and Figure~\ref{fig: beam_search} demonstrate the process of beam search.
The key idea is that we iteratively create new accelerators with some resource and workloads, and at the same time, the unallocated workloads and resources compose a synthetic accelerator (denoted as \texttt{remain\_acc}) to evaluate the utilization of workloads that remain unassigned.
As shown in Algorithm 1, each iteration begins with a parent design point some resources (\texttt{r}) and layers (\texttt{l}) already assigned to a sequence of accelerators(\texttt{accs})(line 5-6).
Then we search all possible resource and workload to create a new accelerator, and construct a new \texttt{remain\_acc} (lines 7-10).
During the search process, we allocate various kinds of resources proportionally.
The \texttt{create\_acc} step uses the same brute-force search approach method as discussed in Figure~\ref{fig: design_space_exploration}.
Only after creating the accelerators, can its utilization be obtained, we then prune this child if its utilization $>$ 1 (line 11).
As the utilization and pipeline constraints are followed when creating new accelerators, once the utilization of the \texttt{remain\_acc} is $\leq$ 1, we can turn this synthetic accelerator into a real one and get a feasible solution (lines 13-14).
If the utilization exceeds 1, this does not indicate failure.
Instead, the design point is also retained in the children, and the algorithm proceeds to further partitioning in subsequent iterations (line 12).
At the end of each iteration, we keep the top $B$ designs in terms of maximum utilization across all accelerators, then use them as parents in the next iteration and repeat this process until the maximum accelerator number is reached.
The beam search can search solutions with multiple $M$ in one run.

The proposed beam search design can generate near-optimal results since our proposed metric, $max(Util)$ across all accelerators, can represent the level of hardware efficiency, which is improved via proper resource and workload partitions.
As shown in Figure~\ref{fig: beam_search}, during the growth of one parent design, all possible partition strategies will be searched.
(1) If too much workload is assigned yet too few resources are allocated to the new accelerator, it is impossible to achieve utilization $<$ 1, this child will be pruned directly (Figure~\ref{fig: beam_search}(B)).
(2) If too little workload is assigned while too many resources are allocated, though the new accelerator will have small utilization, the \texttt{remain\_acc} will have an increased utilization as the lack of computational resources. 
As a result, this child will have a high $max(Util)$ and will not be selected for the next iteration.
(3) For children with inefficient partition, due to the shape mismatch, the new accelerator will have low performance of some (or all) the layers executing on it, thus it either has a high utilization, or occupies more resources.
As a result, these designs will be outperformed by the efficiently partitioned ones.(Figure~\ref{fig: beam_search}(C, D)).

\noindent\textbf{Complexity analysis:} 
For the baseline brute-force search, the complexity can be represented as $O(M\cdot\binom{M}{R+1}\cdot \prod_{i=1}^{n}{\binom{M}{L_i+1}})$, where the complexity within \texttt{create\_acc} is omitted, as it searches a fixed set of parameters and the complexity is constant. 
For the beam search, at most $(\mbox(max\_M)-2)\cdot B + 1$ parent are generated.
For each parent, \texttt{create\_acc} is used at most $R\cdot\prod_{i=1}^{n}{L_i}\cdot2$ times.
As a result, the complexity of beam search is $O(\mbox{max\_M}\cdot B\cdot R\cdot\prod_{i=1}^{n}{L_i})$.

\vspace{-10pt}
\section{Evaluation}
\label{sec: evaluation}


\subsection{Experiment Setup}
We prototype the design searched by our proposed Beam search algorithm on the Versal VCK5000 platform with customized dataflow and control logic adapted for the preemption mechanism and FIFO/EDF scheduling. 
During DSE, we leverage the performance model in CHARM~\cite{zhuang2023charm} to estimate the operation latency for Equations~\ref{eq: acc_execution_length} and~\ref{eq: acc_preemption_overhead}. Notably, the original CHARM DSE is not real-time aware, so these estimates are integrated into our real-time-oriented search. 
The search result for hardware configuration is passed to the CHARM code generator for the accelerator design.
Finally, we customize the generated code for the preemption support and control flows.

We evaluate workloads extracted from real-world applications in both point cloud processing and image processing domains.
We select five applications and truncate their workloads to speed up DSE: PointNet~\cite{qi2017pointnet} (full model), Point Transformer~\cite{wu2024point} (2 blocks), MLP-Mixer~\cite{tolstikhin2021mlp} (2 blocks), Res-MLP~\cite{touvron2022resmlp} (4 blocks), and DeiT-T~\cite{touvron2021training} (2 blocks). Within each model, blocks are the same, but the layers within a block differ; thus, these truncations preserve the heterogeneity of the layers.
On the single accelerator design, these workloads reach 0.23 ms, 0.99ms, 0.30ms, 0.38ms, and 0.14ms, respectively. 
We denote them as \textbf{P'} and use them as a reference for period generation.
To generate various tasksets, we choose one application from the point cloud processing domain (i.e., PointNet or Point Transformer) and one from the image processing domain.
Different periods are then assigned to each task.
\vspace{-10pt}
\subsection{SRT-Schedulability and Utilization}

We first compare SRT-guided (SG) DSE  ~vs. throughput-guided (TG) DSE  on SRT-schedulability and utilizations. 
For the \M~SRT-guided beam search, we set the \texttt{max\_M} to 4 and \texttt{B} to 8.
The value of \texttt{max\_M} is determined through empirical evaluation. Across a broad range of experiments, we observe that most of the feasible design points and the design with the best maximum utilization are found when M $\leq$ 4.
We use CHARM DSE as the throughput-guided baseline.
To generate different tasksets with the same application combination, 
we set the period of each task via switching the $\textbf{P'}/P_1$ and $\textbf{P'}/P_2$ ratio; a larger ratio means a smaller period and heavier workload.
For each DSE result, FIFO and EDF scheduling are used.
Specifically, as CHARM does not follow the pipelined topology constraint, we need to extend the definition of FIFO to (1) baseline FIFO w/o polling, where for job segment $\tau^s_{i,j}$ on $acc^k$, it can be ready only if all segments of the previous job $\tau_{i,j-1}$ on $acc^k$ are finished. This scheduling is introduced in~\cite{dong2022schedulability}.
(2) FIFO w/t polling, where for job segment $\tau^s_{i,j}$ on $acc^k$, it can be ready as long as its corresponding segment in the previous job, $\tau^s_{i,j-1}$, is finished. 
This scheduling is not supported by existing implementations and can be achieved by extending our control flow and allowing the schedulers to release to each other.
For different scheduling methods, the SG+FIFO is guaranteed to be schedulable as long as $max(util)<1$ according to the real-time theory, whereas SG+EDF does not provide the same schedulability guarantee due to the preemption overhead introduced.
Neither are the TG designs, as no real-time theory is used to guarantee the schedulability.
For these designs, we conduct simulations for >100$\times$ period. If an accumulation of unprocessed jobs is detected, this design is non-schedulable.

As shown in Figure~\ref{fig: success_rate}, \M~reaches the largest area of SRT-schedulable among all the designs tested.
Comparing different scheduling methods, the SG+EDF design reaches the same success numbers as the SG+FIFO, but TG+EDF performs worse than TG+FIFO w/t polling.
This is because in TG+EDF, the preemption happens frequently, introducing much overhead that affects the schedulability.
However, in SG+EDF, as the pipelined topology is applied, the preemption is 10$\times$ less likely to happen than TG+EDF, according to the simulation results. 
The overhead is thus much smaller.
For the FIFO designs, FIFO w/o polling performs worse as the newly-released jobs are frequently blocked by the old ones, even when the accelerator is idle.
In summary, the SRT-guided designs achieve 2.28$\times$, 2.28$\times$, 1.44$\times$, 2.25$\times$, 2.17$\times$, 2.10$\times$ more feasible solutions compared with the best baseline (TG+FIFO w/t polling).

\begin{figure}
    \centering
    \includegraphics[width=0.85\linewidth]{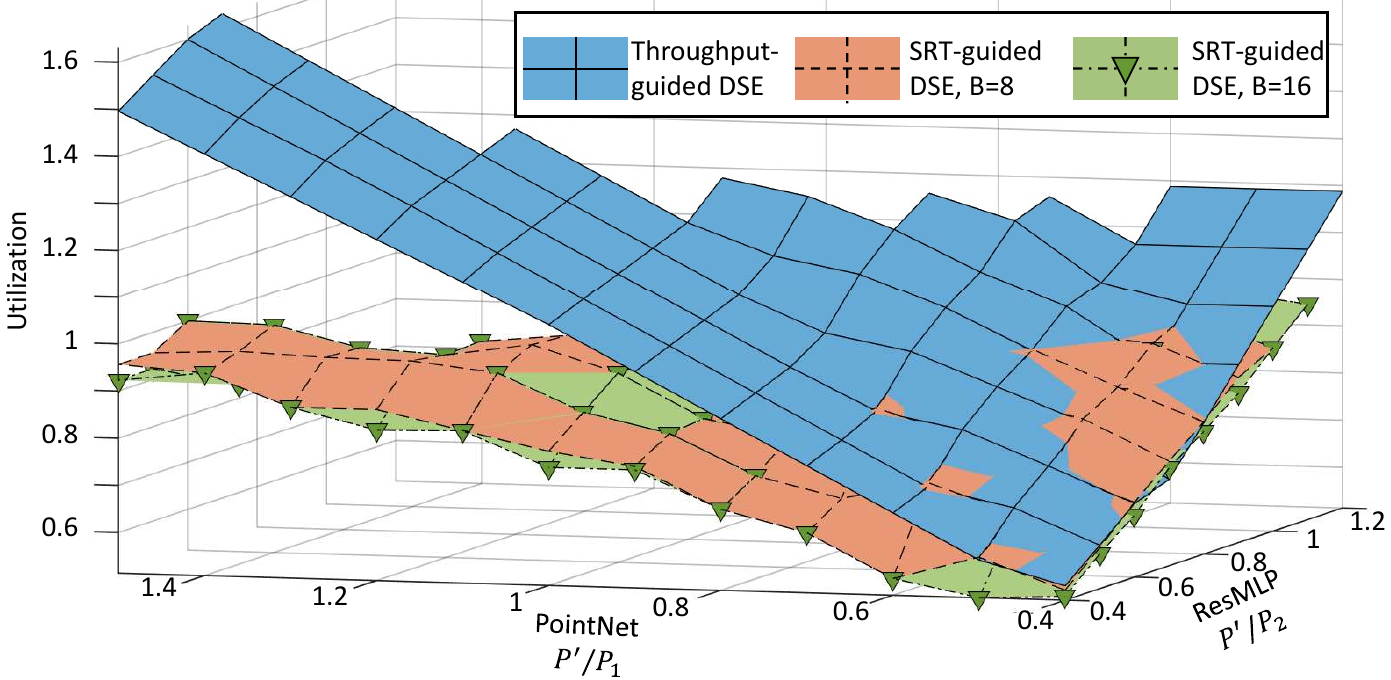}
    \vspace{-12pt}
    \caption{Accelerator maximum utilization on various period setups of PointNet-ResMLP combination.}
    \label{fig: util_distribution}
    \vspace{-10pt}
\end{figure}
\M~also optimizes for the total utilization metric. 
Figure~\ref{fig: util_distribution} demonstrates the utilization distribution with different period setups of the PointNet-ResMLP combination.
The throughput-guided DSE generates higher $max(util)$ than the SRT-guided ones.
with a similar amount of workload, the throughput-guided designs are slightly better when $P_0\simeq P_1$(in the middle) than when there is a large difference(close to x- and y-axis). 
This is because the throughput-guided designs are unaware of the periods of the taskset.
When such similar periods occur, the throughput-guided DSE may generate better results than the SRT-guided design with small beam size, as some global optimal points may be ignored.
When expanding $B$ to 16, the SRT-guided designs are better at every point.
Across different application combinations, \M~has 3.7\%, 4.6\%, -2.4\%, 6.2\%,  3.9\%, 5.1\% better average utilization than the throughput-guided DSE.
Here in the PointNet+DeiT-T group, \M~performs worse also because of the suboptimality. 
When increasing B to 16 and 32, \M~will be 0.07\% and 3.5\% better.

\vspace{-5pt}
\subsection{Response Time Analysis}
\begin{figure}
    \centering
    \includegraphics[width=0.85\linewidth]{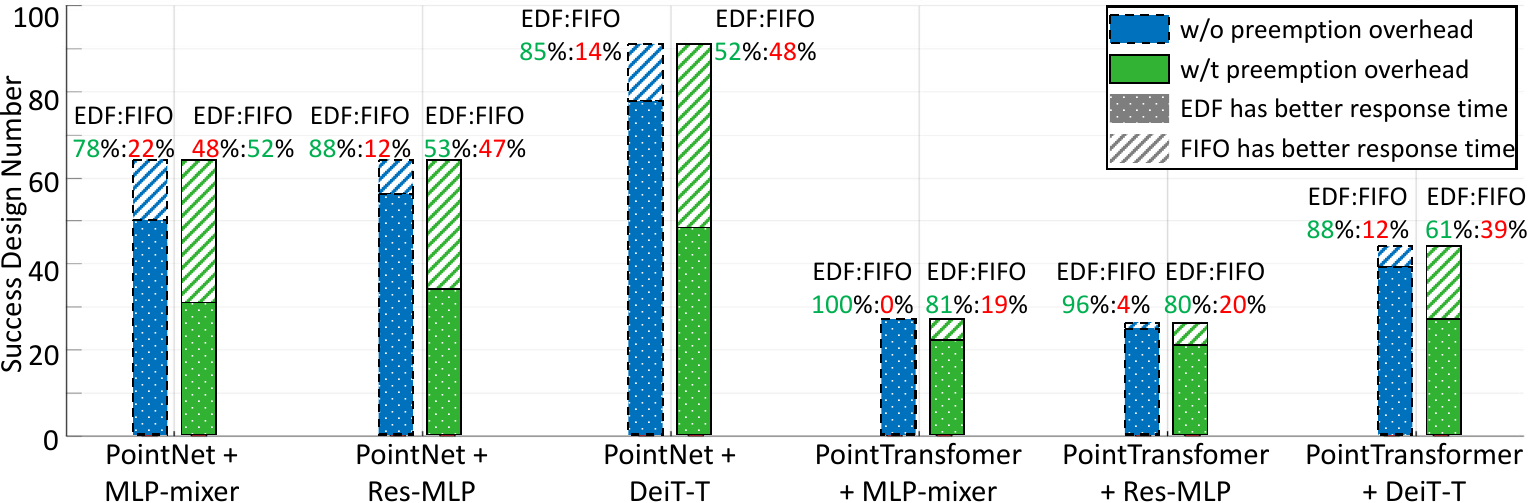}
    \vspace{-12pt}
    \caption{The response time statistics of FIFO vs. EDF scheduling on various application combinations.}
    \label{fig: resp_time_analysis}
    \vspace{-15pt}
\end{figure}


We conduct the response time analysis to compare the performance of the FIFO and EDF scheduling method introduced by \M~as shown in  Figure~\ref{fig: resp_time_analysis}.
Two sets of simulations with and without considering the preemption overhead are conducted on the SRT-guided DSE result designs.
Without considering the overhead, EDF scheduling will have a better response time in most cases.
However, the overhead is inevitable. 
When considering overhead in scheduling, the rate of tasksets that perform better with EDF drops.
Still, in the Point Transformer-related groups, EDF remains better in 81\%, 80\%, and 61\% designs. 
This is because Point Transformer has a much larger execution time, indicated by the P' demonstrated above.
Without EDF, the other workload with smaller execution time and period will be blocked by this large task, resulting in larger response times.
Our analysis shows that there is no dominantly better scheduling method in \M~SRT system: the FIFO scheduling may incur larger response time when workloads vary largely, whereas the EDF introduces overhead and may not guarantee schedulability.
\M~has the flexibility of offering both.

\vspace{-5pt}
\subsection{Beam Search Quality}
\begin{figure}
    \centering
    \includegraphics[width=0.85\linewidth]{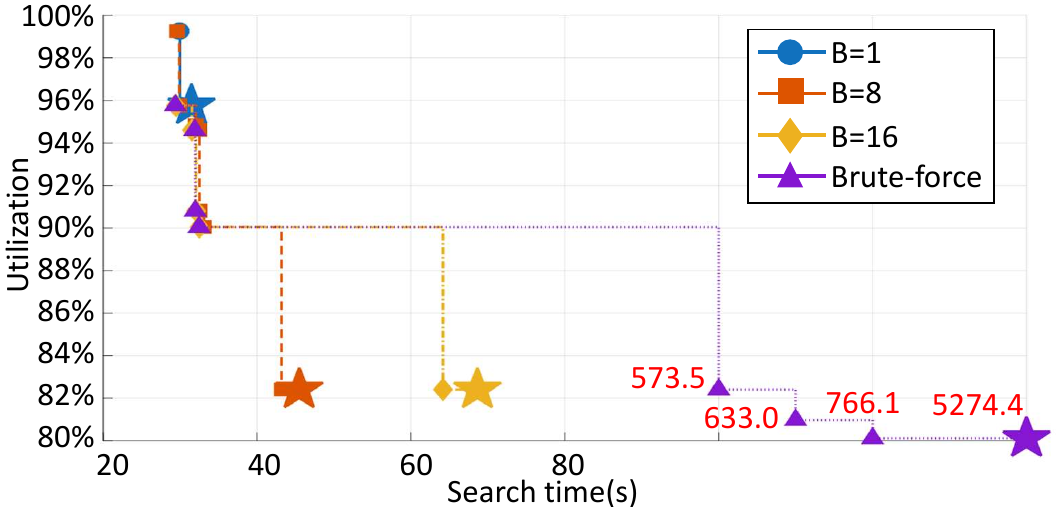}
    \vspace{-12pt}
    \caption{Search time comparison of various beam search setups and brute-force search of PointNet-Deit-T combination.}
    \vspace{-15pt}
    \label{fig: quality}
\end{figure}

We demonstrate the beam search quality via Figure~\ref{fig: quality}, where we compare different beam search and the brute-force search implemented via BFS, where B is set to +inf.
All DSE is conducted on a server with one AMD Ryzen 9965WX CPU.
Comparing various beam searches, the B=1 search finishes \jz{with the least search time}, but generates sub-optimal results. 
As B increases, more candidates are tested, and the result reaches near-optimal.
For all the designs, the first feasible solution is generated in a small and similar time, \jz{since the number of candidates in the early iterations is very similar}.
However, in the brute-force search, the number of candidates will increase exponentially; as a result, it takes >500 seconds to find the same solution as beam search and > 5000 seconds to finish the brute-force search, which are 13.3$\times$ and 117.2$\times$ longer than the B=8 beam search, with 2.3\% improvement in best $max(util)$ searched.

\vspace{-7pt}
\subsection{Discussion}
\label{sec: discussion}
\noindent\textbf{Reasons to introduce real-time theory and pipeline constraint:}
Real-time theory is an important guideline for accelerator design; \jz{Only with real-time theory can we mathematically guarantee deadline compliance.}
\M~conducts DSE under this assumption. We leave more advanced real-time theory on more complicated accelerators and the paired accelerator designs for future discussions.

\noindent\textbf{The impact of pipeline constraint:}
The limitations of pipeline constraint are twofold:
(1) The constraint requires the application to be a sequence of layers.
\jz{For more complicated topologies, such as transformer blocks, layer-level parallelism is not exploited; we leave this extension for future work.}
Instead, \M~explores the parallelism from different jobs of one task, and different tasks.
(2) When a task has a large variance in layer shapes that are interleaved, the DSE performance will be limited due to hardware inefficiency. However, with the pipeline constraint, \M~can always guarantee schedulability for designs satisfying the utilization constraints.


\smallskip
{\small
{\noindent\textbf{ACKNOWLEDGEMENTS --}} This work is supported in part by Brown University New Faculty Start-up Grant, NSF awards 
\#2140346, 
\#2231523, 
\#2348306, 
\#2441179, 
\#2511445, 
\#2518375, 
\#2536952. 
We thank AMD for the hardware and software donations.
\vspace{-10pt}
\bibliographystyle{unsrt}
\balance
\bibliography{ref_filtered}

@INPROCEEDINGS{ghodrati2020planaria,
  author={Ghodrati, Soroush and Ahn, Byung Hoon and Kyung Kim, Joon and Kinzer, Sean and Yatham, Brahmendra Reddy and Alla, Navateja and Sharma, Hardik and Alian, Mohammad and Ebrahimi, Eiman and Kim, Nam Sung and Young, Cliff and Esmaeilzadeh, Hadi},
  booktitle={2020 53rd Annual IEEE/ACM International Symposium on Microarchitecture (MICRO)}, 
  title={{Planaria: Dynamic Architecture Fission for Spatial Multi-Tenant Acceleration of Deep Neural Networks}}, 
  year={2020},
  volume={},
  number={},
  pages={681-697},
  doi={10.1109/MICRO50266.2020.00062}}

@misc{vitisai,
  author       = {{AMD}},
  title = {{AMD Vitis™ AI Software}},
  url          = {https://www.amd.com/en/products/software/vitis-ai.html}
}

@inproceedings{basalama2025streamhls,
  title = {{Stream-HLS: Towards Automatic Dataflow Acceleration}},
  author={Basalama, Suhail and Cong, Jason},
  booktitle = {{Proceedings of the 2025 ACM/SIGDA International Symposium on Field Programmable Gate Arrays}},
  pages={103--114},
  year={2025}
}

@inproceedings{Dream,
  title = {{DREAM: A Dynamic Scheduler for Dynamic Real-time Multi-model ML Workloads}},
  author={Kim, Seah and Kwon, Hyoukjun and Song, Jinook and Jo, Jihyuck and Chen, Yu-Hsin and Lai, Liangzhen and Chandra, Vikas},
  booktitle = {{Proceedings of the 28th ACM International Conference on Architectural Support for Programming Languages and Operating Systems, Volume 4}},
  pages={73--86},
  year={2023}
}

@inproceedings{dong2017optimal,
  author       = {Zheng Dong and
                  Cong Liu and
                  Alan Gatherer and
                  Lee McFearin and
                  Peter Yan and
                  James H. Anderson},
  editor       = {Marko Bertogna},
  title        = {{Optimal Dataflow Scheduling on a Heterogeneous Multiprocessor With
                  Reduced Response Time Bounds}},
  booktitle    = {29th Euromicro Conference on Real-Time Systems, {ECRTS} 2017, Dubrovnik,
                  Croatia, June 27-30, 2017},
  series       = {LIPIcs},
  pages        = {15:1--15:22},
  publisher    = {Schloss Dagstuhl - Leibniz-Zentrum f{\"{u}}r Informatik},
  year         = {2017},
  url          = {https://doi.org/10.4230/LIPIcs.ECRTS.2017.15},
  doi          = {10.4230/LIPICS.ECRTS.2017.15},
  bibsource    = {dblp computer science bibliography, https://dblp.org}
}

@inproceedings{devi2005tardiness,
author = {Devi, UmaMaheswari C. and Anderson, James H.},
title = {{Tardiness Bounds under Global EDF Scheduling on a Multiprocessor}},
year = {2005},
isbn = {0769524907},
publisher = {IEEE Computer Society},
address = {USA},
url = {https://doi.org/10.1109/RTSS.2005.39},
doi = {10.1109/RTSS.2005.39},
booktitle = {Proceedings of the 26th IEEE International Real-Time Systems Symposium},
pages = {330–341},
numpages = {12},
series = {RTSS '05}
}

@inproceedings{zhang2020dnnexplorer,
author = {Zhang, Xiaofan and Ye, Hanchen and Wang, Junsong and Lin, Yonghua and Xiong, Jinjun and Hwu, Wen-mei and Chen, Deming},
title = {{DNNExplorer: a framework for modeling and exploring a novel paradigm of FPGA-based DNN accelerator}},
year = {2020},
isbn = {9781450380263},
publisher = {Association for Computing Machinery},
address = {New York, NY, USA},
url = {https://doi.org/10.1145/3400302.3415609},
doi = {10.1145/3400302.3415609},
booktitle = {Proceedings of the 39th International Conference on Computer-Aided Design},
articleno = {61},
numpages = {9},
location = {Virtual Event, USA},
series = {ICCAD '20}
}

@ARTICLE{dong2022schedulability,
  author={Dong, Zheng and Liu, Cong},
  journal={IEEE Transactions on Computer-Aided Design of Integrated Circuits and Systems}, 
  title={{Schedulability Analysis for Coscheduling Real-Time Tasks on Multiprocessors}}, 
  year={2022},
  volume={41},
  number={11},
  pages={4721-4732},
  doi={10.1109/TCAD.2022.3141971}}

@article{dally2020domain,
author = {Dally, William J. and Turakhia, Yatish and Han, Song},
title = {{Domain-specific hardware accelerators}},
year = {2020},
issue_date = {July 2020},
publisher = {Association for Computing Machinery},
address = {New York, NY, USA},
volume = {63},
number = {7},
issn = {0001-0782},
url = {https://doi.org/10.1145/3361682},
doi = {10.1145/3361682},
journal = {Commun. ACM},
month = jun,
pages = {48–57},
numpages = {10}
}

@ARTICLE{eq_vit_tcad,
  author={Dong, Peiyan and Zhuang, Jinming and Yang, Zhuoping and Ji, Shixin and Li, Yanyu and Xu, Dongkuan and Huang, Heng and Hu, Jingtong and Jones, Alex K. and Shi, Yiyu and Wang, Yanzhi and Zhou, Peipei},
  journal={IEEE Transactions on Computer-Aided Design of Integrated Circuits and Systems}, 
  title={{EQ-ViT: Algorithm-Hardware Co-Design for End-to-End Acceleration of Real-Time Vision Transformer Inference on Versal ACAP Architecture}}, 
  year={2024},
  volume={43},
  number={11},
  pages={3949-3960},
  doi={10.1109/TCAD.2024.3443692}}

@inproceedings{ji2025art,
author = {Ji, Shixin and Chen, Xingzhen and Zhuang, Jinming and Zhang, Wei and Yang, Zhuoping and Schultz, Sarah and Song, Yukai and Hu, Jingtong and Jones, Alex and Dong, Zheng and Zhou, Peipei},
title = {{ART: Customizing Accelerators for DNN-Enabled Real-Time Safety-Critical Systems}},
year = {2025},
isbn = {9798400714962},
publisher = {Association for Computing Machinery},
address = {New York, NY, USA},
url = {https://doi.org/10.1145/3716368.3735215},
doi = {10.1145/3716368.3735215},
booktitle = {Proceedings of the Great Lakes Symposium on VLSI 2025},
pages = {442–449},
numpages = {8},
location = {
},
series = {GLSVLSI '25}
}

@inproceedings{yang2023aim,
  title = {{AIM: Accelerating Arbitrary-precision Integer Multiplication on Heterogeneous Reconfigurable Computing Platform Versal ACAP}},
  author={Yang, Zhuoping and Zhuang, Jinming and Yin, Jiaqi and Yu, Cunxi and Jones, Alex K and Zhou, Peipei},
  booktitle = {{2023 IEEE/ACM International Conference on Computer Aided Design (ICCAD)}},
  pages={1--9},
  year={2023},
  organization={IEEE}
}

@INPROCEEDINGS{kim2023moca,
  author={Kim, Seah and Genc, Hasan and Nikiforov, Vadim Vadimovich and Asanović, Krste and Nikolić, Borivoje and Shao, Yakun Sophia},
  booktitle={2023 IEEE International Symposium on High-Performance Computer Architecture (HPCA)}, 
  title={{MoCA: Memory-Centric, Adaptive Execution for Multi-Tenant Deep Neural Networks}}, 
  year={2023},
  volume={},
  number={},
  pages={828-841},
  doi={10.1109/HPCA56546.2023.10071035}}

@INPROCEEDINGS{DERCA,
  author={Ji, Shixin and Yang, Zhuoping and Chen, Xingzhen and Zhang, Wei and Zhuang, Jinming and Jones, Alex K. and Dong, Zheng and Zhou, Peipei},
  booktitle = {{2025 IEEE Real-Time Systems Symposium (RTSS)}}, 
  title = {{DERCA: DetERministic Cycle-Level Accelerator on Reconfigurable Platforms in DNN-Enabled Real-Time Safety-Critical Systems}}, 
  year={2025},
  volume={},
  number={},
  pages={392-405},
  doi={10.1109/RTSS66672.2025.00039}}

@inproceedings{ji2025towards,
author = {Ji, Shixin and Chen, Xingzhen and Zhang, Wei and Yang, Zhuoping and Zhuang, Jinming and Schultz, Sarah and Song, Yukai and Hu, Jingtong and Jones, Alex K. and Dong, Zheng and Zhou, Peipei},
title = {{Towards Accelerator Customization in Real-time Safety-critical Systems}},
year = {2025},
isbn = {9798400713965},
publisher = {Association for Computing Machinery},
address = {New York, NY, USA},
url = {https://doi.org/10.1145/3706628.3708841},
doi = {10.1145/3706628.3708841},
booktitle = {Proceedings of the 2025 ACM/SIGDA International Symposium on Field Programmable Gate Arrays},
pages = {181},
numpages = {1},
location = {Monterey, CA, USA},
series = {FPGA '25}
}

@inproceedings{zhuang2023charm,
author = {Zhuang, Jinming and Lau, Jason and Ye, Hanchen and Yang, Zhuoping and Du, Yubo and Lo, Jack and Denolf, Kristof and Neuendorffer, Stephen and Jones, Alex and Hu, Jingtong and Chen, Deming and Cong, Jason and Zhou, Peipei},
title = {{CHARM: Composing Heterogeneous AcceleRators for Matrix Multiply on Versal ACAP Architecture}},
year = {2023},
isbn = {9781450394178},
publisher = {Association for Computing Machinery},
address = {New York, NY, USA},
url = {https://doi.org/10.1145/3543622.3573210},
doi = {10.1145/3543622.3573210},
booktitle = {Proceedings of the 2023 ACM/SIGDA International Symposium on Field Programmable Gate Arrays},
pages = {153–164},
numpages = {12},
location = {Monterey, CA, USA},
series = {FPGA '23}
}

@inproceedings{wu2024point,
  title = {{Point transformer v3: Simpler faster stronger}},
  author={Wu, Xiaoyang and Jiang, Li and Wang, Peng-Shuai and Liu, Zhijian and Liu, Xihui and Qiao, Yu and Ouyang, Wanli and He, Tong and Zhao, Hengshuang},
  booktitle = {{Proceedings of the IEEE/CVF conference on computer vision and pattern recognition}},
  pages={4840--4851},
  year={2024}
}

@inproceedings{jiang2023blueface,
  title = {{BlueFace: Integrating an Accelerator into the Core’s Pipeline through Algorithm-Interface Co-Design for Real-Time SoCs}},
  author={Jiang, Zhe and Fisher, Nathan and Guan, Nan and Dong, Zheng},
  booktitle = {{2023 60th ACM/IEEE Design Automation Conference (DAC)}},
  pages={1--6},
  year={2023},
  organization={IEEE}
}

@inproceedings{gao2023layer,
  title = {{Layer-Puzzle: Allocating and Scheduling Multi-task on Multi-core NPUs by Using Layer Heterogeneity}},
  author={Gao, Chengsi and Wang, Ying and Liu, Cheng and Wang, Mengdi and Chen, Weiwei and Han, Yinhe and Zhang, Lei},
  booktitle = {{2023 Design, Automation \& Test in Europe Conference \& Exhibition (DATE)}},
  pages={1--6},
  year={2023},
  organization={IEEE}
}

@INPROCEEDINGS{kwon2021herald,
  author={Kwon, Hyoukjun and Lai, Liangzhen and Pellauer, Michael and Krishna, Tushar and Chen, Yu-Hsin and Chandra, Vikas},
  booktitle={2021 IEEE International Symposium on High-Performance Computer Architecture (HPCA)}, 
  title={{Heterogeneous Dataflow Accelerators for Multi-DNN Workloads}}, 
  year={2021},
  volume={},
  number={},
  pages={71-83},
  doi={10.1109/HPCA51647.2021.00016}}

@inproceedings{cai2023set,
author = {Cai, Jingwei and Wei, Yuchen and Wu, Zuotong and Peng, Sen and Ma, Kaisheng},
title = {{Inter-layer Scheduling Space Definition and Exploration for Tiled Accelerators}},
year = {2023},
isbn = {9798400700958},
publisher = {Association for Computing Machinery},
address = {New York, NY, USA},
url = {https://doi.org/10.1145/3579371.3589048},
doi = {10.1145/3579371.3589048},
booktitle = {Proceedings of the 50th Annual International Symposium on Computer Architecture},
articleno = {13},
numpages = {17},
location = {Orlando, FL, USA},
series = {ISCA '23}
}

@INPROCEEDINGS{oh2021layerweaver,
  author={Oh, Young H. and Kim, Seonghak and Jin, Yunho and Son, Sam and Bae, Jonghyun and Lee, Jongsung and Park, Yeonhong and Kim, Dong Uk and Ham, Tae Jun and Lee, Jae W.},
  booktitle={2021 IEEE International Symposium on High-Performance Computer Architecture (HPCA)}, 
  title={{Layerweaver: Maximizing Resource Utilization of Neural Processing Units via Layer-Wise Scheduling}}, 
  year={2021},
  volume={},
  number={},
  pages={584-597},
  doi={10.1109/HPCA51647.2021.00056}}

@inproceedings{ssr,
author = {Zhuang, Jinming and Yang, Zhuoping and Ji, Shixin and Huang, Heng and Jones, Alex K. and Hu, Jingtong and Shi, Yiyu and Zhou, Peipei},
title = {{SSR: Spatial Sequential Hybrid Architecture for Latency Throughput Tradeoff in Transformer Acceleration}},
year = {2024},
isbn = {9798400704185},
publisher = {Association for Computing Machinery},
address = {New York, NY, USA},
url = {https://doi.org/10.1145/3626202.3637569},
doi = {10.1145/3626202.3637569},
booktitle = {{Proceedings of the 2024 ACM/SIGDA International Symposium on Field Programmable Gate Arrays}},
pages = {55–66},
numpages = {12},
location = {Monterey, CA, USA},
series = {FPGA '24}
}

@INPROCEEDINGS{choi2020prema,
  author={Choi, Yujeong and Rhu, Minsoo},
  booktitle={2020 IEEE International Symposium on High Performance Computer Architecture (HPCA)}, 
  title={{PREMA: A Predictive Multi-Task Scheduling Algorithm For Preemptible Neural Processing Units}}, 
  year={2020},
  volume={},
  number={},
  pages={220-233},
  doi={10.1109/HPCA47549.2020.00027}}

@INPROCEEDINGS{restuccia2021time,
  author={Restuccia, Francesco and Biondi, Alessandro},
  booktitle={2021 IEEE Real-Time Systems Symposium (RTSS)}, 
  title={{Time-Predictable Acceleration of Deep Neural Networks on FPGA SoC Platforms}}, 
  year={2021},
  volume={},
  number={},
  pages={441-454},
  doi={10.1109/RTSS52674.2021.00047}}

@InProceedings{touvron2021training,
  title = 	 {{Training data-efficient image transformers \& distillation through attention}},
  author =       {Touvron, Hugo and Cord, Matthieu and Douze, Matthijs and Massa, Francisco and Sablayrolles, Alexandre and Jegou, Herve},
  booktitle = 	 {Proceedings of the 38th International Conference on Machine Learning},
  pages = 	 {10347--10357},
  year = 	 {2021},
  editor = 	 {Meila, Marina and Zhang, Tong},
  volume = 	 {139},
  series = 	 {Proceedings of Machine Learning Research},
  month = 	 {18--24 Jul},
  publisher =    {PMLR},
  url = 	 {https://proceedings.mlr.press/v139/touvron21a.html},
}

@ARTICLE{touvron2022resmlp,
  author={Touvron, Hugo and Bojanowski, Piotr and Caron, Mathilde and Cord, Matthieu and El-Nouby, Alaaeldin and Grave, Edouard and Izacard, Gautier and Joulin, Armand and Synnaeve, Gabriel and Verbeek, Jakob and Jégou, Hervé},
  journal={IEEE Transactions on Pattern Analysis and Machine Intelligence}, 
  title={{ResMLP: Feedforward Networks for Image Classification With Data-Efficient Training}}, 
  year={2023},
  volume={45},
  number={4},
  pages={5314-5321},
  doi={10.1109/TPAMI.2022.3206148}}

@ARTICLE{zeng2022serving,
  author={Zeng, Shulin and Dai, Guohao and Zhang, Niansong and Yang, Xinhao and Zhang, Haoyu and Zhu, Zhenhua and Yang, Huazhong and Wang, Yu},
  journal={IEEE Transactions on Computers}, 
  title={{Serving Multi-DNN Workloads on FPGAs: A Coordinated Architecture, Scheduling, and Mapping Perspective}}, 
  year={2023},
  volume={72},
  number={5},
  pages={1314-1328},
  doi={10.1109/TC.2022.3214113}}

@InProceedings{qi2017pointnet,
author = {Qi, Charles R. and Su, Hao and Mo, Kaichun and Guibas, Leonidas J.},
title = {{PointNet: Deep Learning on Point Sets for 3D Classification and Segmentation}},
booktitle = {Proceedings of the IEEE Conference on Computer Vision and Pattern Recognition (CVPR)},
month = {July},
year = {2017}
}

@ARTICLE{wang2023cd,
  author={Wang, Chunyang and Bai, Yuebin and Sun, Desen},
  journal={IEEE Transactions on Parallel and Distributed Systems}, 
  title={{CD-MSA: Cooperative and Deadline-Aware Scheduling for Efficient Multi-Tenancy on DNN Accelerators}}, 
  year={2023},
  volume={34},
  number={7},
  pages={2091-2106},
  doi={10.1109/TPDS.2023.3276759}}

@article{li2023fixed,
  title = {{Fixed-point FPGA Implementation of the FFT Accumulation Method for Real-time Cyclostationary Analysis}},
  author={Li, Carol Jingyi and Li, Xiangwei and Lou, Binglei and Jin, Craig T and Boland, David and Leong, Philip HW},
  journal={ACM Transactions on Reconfigurable Technology and Systems},
  volume={16},
  number={3},
  pages={1--28},
  year={2023},
  publisher={ACM New York, NY}
}

@article{liu1973scheduling,
author = {Liu, C. L. and Layland, James W.},
title = {{Scheduling Algorithms for Multiprogramming in a Hard-Real-Time Environment}},
year = {1973},
issue_date = {Jan. 1973},
publisher = {Association for Computing Machinery},
address = {New York, NY, USA},
volume = {20},
number = {1},
issn = {0004-5411},
url = {https://doi.org/10.1145/321738.321743},
doi = {10.1145/321738.321743},
journal = {J. ACM},
month = jan,
pages = {46–61},
numpages = {16}
}

@inproceedings{guan2024mesc,
  title = {{MESC: Re-thinking Algorithmic Priority and/or Criticality Inversions for Heterogeneous MCSs}},
  author={Guan, Jiapeng and Wei, Ran and You, Dean and Wang, Yingquan and Yang, Ruizhe and Wang, Hui and Jiang, Zhe},
  booktitle = {{2024 IEEE Real-Time Systems Symposium (RTSS)}},
  pages={1--14},
  year={2024},
  organization={IEEE}
}

@inproceedings{tolstikhin2021mlp,
 author = {Tolstikhin, Ilya O and Houlsby, Neil and Kolesnikov, Alexander and Beyer, Lucas and Zhai, Xiaohua and Unterthiner, Thomas and Yung, Jessica and Steiner, Andreas and Keysers, Daniel and Uszkoreit, Jakob and Lucic, Mario and Dosovitskiy, Alexey},
 booktitle = {Advances in Neural Information Processing Systems},
 editor = {M. Ranzato and A. Beygelzimer and Y. Dauphin and P.S. Liang and J. Wortman Vaughan},
 pages = {24261--24272},
 publisher = {Curran Associates, Inc.},
 title = {{MLP-Mixer: An all-MLP Architecture for Vision}},
 url = {https://proceedings.neurips.cc/paper_files/paper/2021/file/cba0a4ee5ccd02fda0fe3f9a3e7b89fe-Paper.pdf},
 volume = {34},
 year = {2021}
}

@inproceedings{fpga25aries,
author = {Zhuang, Jinming and Xiang, Shaojie and Chen, Hongzheng and Zhang, Niansong and Yang, Zhuoping and Mao, Tony and Zhang, Zhiru and Zhou, Peipei},
title = {{ARIES: An Agile MLIR-Based Compilation Flow for Reconfigurable Devices with AI Engines}},
year = {2024},
booktitle = {{Proceedings of the 2025 ACM/SIGDA International Symposium on Field Programmable Gate Arrays}},
series = {FPGA '25}
}

\end{document}